\documentstyle[prl,preprint,aps]{revtex}
\begin{document}
\tolerance=800
\draft
\newcommand{\lan}{$\mbox{La}_{2-x}\mbox{Sr}_x\mbox{CuO}_4 \;$}
\newcommand{\vk}{{\bf k}}
\newcommand{\vp}{{\bf p}}
\newcommand{\vq}{{\bf q}}
\newcommand{\va}{{\bf a}}
\newcommand{\bo}{\tilde{\omega}}
\newcommand{\vf}{\frac{|\vp+\vq|^2}{m_F^2}}
\newcommand{\vb}{\frac{|\vq|^2}{m_B^2}}
\def\vx{{\bf x}}
\def\vy{{\bf y}}
\newcommand{\ba}{\begin{eqnarray}}
\newcommand{\ea}{\end{eqnarray}}
\newcommand{\be}{\begin{equation}}
\newcommand{\ee}{\end{equation}}
\title{One-Electron Spectral Weight of Doped Mott Insulators \\
 Gauge Field Theory Approach}
\author{H. C. Lee}
\address{
James Franck Institute and  the Department of Physics \\
at the University of Chicago, Chicago, IL 60637 \\
{\rm e-mail}:\hspace{.1in}{\tt hyunlee@control.uchicago.edu } }

\date{January 3, 1996}
\maketitle
\begin{abstract}
\indent  In this note we study the electron spectral weight of doped Mott insulators  based on the 2D slave boson gauge field theory. The  vertex
correction  with  static gauge field is calculated in the second order  perturbation theory.  The vertex correction is found to be singular at low energy and requires non-perturbative treatments. 
\end{abstract}
\vskip 1.0cm
\pacs{{\rm PACS numbers}: \hspace{.05in} 74.20.Mn,  74.25.-q,  }
\pagebreak
\section{Introduction}
\indent The properties of  Fermi surface of strongly correlated system are quite interesting. The Fermi surface is determined by one-electron spectral weight and the spectral weight can be found experimentally by angle resolved photo-emission spectroscopy \cite{shen,olson}.\\
\indent There have been many experimental stuides on the structure of Fermi surface of high ${\rm T}_c$ compounds and large Fermi surface with an area consistent with band-strucuture calculation  was found \cite{olson}. The electron Fermi surface (EFS) contains $1-\delta$ electrons per site, where $\delta$ is the hole doping concentration.\\
\indent Theoretically the band structure calculation is the first step in determining the shape of Fermi surface and other one-electron  properties. For weakly interacting system the band structure calculation can explain one-electron spectral properties. \\
\indent However, in the strongly correlated system  like large-U Hubbard model or equivalently $t$-$J$ model it is absolutely necessary to take the correlation 
effect into account in computing one-electron spectral weight. \\
\indent  Motivated by high ${\rm T}_c$ compounds, the two dimensional $t$-$J$ model is widely regarded to capture  the essential physics of anomalous normal state properties of cuprates.  One way of studying $t$-$J$ model  is the two dimensional gauge theory.\\
  Large on-site repulsion imposes the constraint that the electron occupation number at each site should be less than 1. The constraint of $t$-$J$ model can be implemented by expressing the electron operator $c_{i \alpha}$ as
the product of fermion (spinon) $f_{i \alpha}$ and slave boson (holon) $b_i$ and by introuducing gauge field\cite{larkin,lee}. The gauge field represents  the fluctuation of phase of order parameter  $\chi_{ij}=\sum_{\alpha}<\,f_{i \alpha}\,f^{\dag}_{j \alpha}\,>\,$ . The spinons and holons are strongly coupled by gauge field.  To understand the one-electron properties  we have to take 
into account the interaction between spinon and holon  via gauge field . \\
\indent Two dimensional gauge theory can explain  many of anomalous normal state transport properties successfully \cite{lee,paul,ioffe}.  One-electron spectral weight was studied on the mean field level \cite{lee}.  The electron Green function is simply a product of  free fermion and boson Green functions in mean field approximation.
Therefore, the propagating character of fermion Green function  
$ < f(x) f^{\dag}(y) > \sim  e^{ [i (|\vx-\vy|-v_F(t_x-t_y) ]}$
is blocked  by  the localized  boson Green function $~<b(y)b^{\dag}(x)>\sim
\frac{T_0}{T}\,\,e^{(-|\vx-\vy|^2 \,m_B T)} \, $. As a result the coherence of fermions is completely lost.\\
 \indent EFS does not exist in the mean field spectral weight \cite{yu}, which indicates the inadequacy of the mean field spectral weight. Obviously the gauge field fluctuations need to be incoporated.   The non-static gauge field
fluctuation gives large correction only to self-energy of spinons. In particular the vertex correction  is only logarithmically singular \cite{lee}. It turns our that the non-static gauge field does not change the overall feature of mean field momentum distribution function \cite{yu}.\\
\indent For  static gauge field the perturbative
calculation was not feasible due to  infrared divergences. The divergence is, of course, spurious and should be cancelled in any gauge invariant quantity.
Using  gauge invariant path integral  methods the relevant relaxation time due to static gauge
field was identified $\tau_{{\rm in}}^{-1}\approx (T/\chi)^{2/3}\,(m_B m_F^2)^{-1/3} \,$ \cite{lee}. However, the detailed functional form of spectral weight could was not obtained. \\
\indent In this paper we carry out the second order perturbative calculation with  static gauge field.  The spurious infrared divergences are shown to be cancelled.
The vertex correction gives  the dominant contribution.  
\be
A(\epsilon,\vp)=A_{0 b}(\epsilon,\vp)\, \Big(1-\frac{\tau_{{\rm in}}^{-3/2}}{|\epsilon-\mu_B-\xi_\vp|\,\sqrt{T}}\Big)+A_{0 f}(\epsilon,\vp)\,\Big(1-\frac{\tau_{{\rm in}}^{-3/2}}{|\epsilon-\mu_B-\xi_\vp| \sqrt{\tilde{\epsilon}}} \Big),
\ee
where $A_{0 b}(\epsilon,\vp)$  and $A_{0 f}(\epsilon,\vp)$ is the boson and fermion part of the mean field spectral weight. The second order perturbative result
is singular at low energy  but the singular behaviour may disappear after proper resummation or non-perturbative treatment. If the gauge field correction is enhanced near $|\epsilon-\mu_B-\xi_\vp|\approx 0$ it contributes to the formation of EFS.

\section{Spectral Weight in decoupling approximation}
In this section we briefly review 2D slave boson gauge field theory and calculate the one-electron spectral weight in decoupling approximation, which means that the vertex correction is ignored \cite{lee,yu}.\\
\indent Strong on-site Coulomb repulsion forbids double
occupations  and imposes the constraint
$\sum_{\alpha}c^{\dag}_{\alpha}(r)c_{\alpha}(r)\le 1
\;$.
The gauge field is a
tool to deal
with this constraint.   The constraint can be
implemented  by representing an electronic operator
$c_{\alpha}(r)$ by the product of a fictitious  spinon
$f_{\alpha}(r)$ and a holon $b^{\dag}(r)$ that keeps
track of vacant sites: 
$\sum_{\alpha}f^{\dag}_{\alpha}(r)f_{\alpha}(r)+
b^{\dag}(r)b(r)=1$. $f_{\alpha}(r)$ is a fermion, while
$b(r)$ is a boson.
An accepted phenomenological model that captures a vector
character of the interaction has the form \cite{larkin,lee,paul,ioffe}.
\be
\begin{array}{lr}
\label{H2D}
H&= \int d^2 r \Bigl [\; \sum_{\alpha}
f^{\dag}_{\alpha}(r)\left(\,-a_0-\mu_F -\frac{1}{2 m_F} (\nabla-i \va)^2 \,
\right ) f_{\alpha}(r)   \\
&+b^{\dag}(r)\,\left(-a_0-\mu_B-\frac{1}{2 m_B}(\nabla-i \va)^2\right)
\, b(r) \Bigr ]
\end{array}
\ee
The bosons are assumed not to condense, so that the  temperature  must be higher than
$T_0$. $T_0$ is the degeneracy temperature of 2D bosons. In free boson approximation it is given by $\,T_0=2\pi \, \delta/m_B \,$, where $\delta$ is the density of bosons which is close to doping concentration. \\
\indent The propagator of the gauge field can be obtained by integrating out high-freqency and momentum spinon states\cite{larkin,lee}. In transverse gauge,
\be
D_{ij}(i\omega,\vk)=\frac{(\delta_{ij}-\frac{k_i\,k_j}{|\vk|^2})}{\chi |\vk|^2+\gamma |\omega|/|\vk|}\equiv (\delta_{ij}-\frac{k_i\,k_j}{|\vk|^2})\,D(i\omega,\vk),\quad i,j=1,2
\ee
For static gauge field, $\,D(\omega=0,\vk)\equiv D(\vk)=[\,\chi |\vk|^2\,]^{-1} \,$.\\
\indent Electron is a gauge invariant object in gauge theory.
In imaginary time the electron Green function is 
$G_{\alpha}(\tau, r)=-<T_\tau \,c_{\alpha}(r,\tau)\,c_{\alpha}^{\dag}(0)>$.
Electron spectral weight is defined by
$A(\epsilon,\vp)=-2\;\mbox{Im}\,G^R(\epsilon,\vp)$. 
 The electron Green function is a {\em two} particle Green function in terms of spinon and holon.

\begin{equation}
\label{eq:spec}
G_{\alpha}(\tau,r)=-<T_\tau \,f_{\alpha}(\tau,r)\,b^{\dag}(\tau,r)\,f_{\alpha}^{\dag}(0)\,b(0)\,>  
\end{equation}  
If the gauge interaction between spinon and holon is ignored the electron Green function can be factorized.
\begin{equation}
G_{\alpha}(\tau, r)=-G_{F,\alpha}(\tau,r)\, G_B(-\tau,-r),
\end{equation}
where $G_{F,\alpha}(\tau,r)=-<T_\tau \,f_{\alpha}(\tau,r)\,f_{\alpha}^{\dag}(0)>,\;\;G_B(-\tau,-r)=-<T_\tau \,b(0)\,b^{\dag}(\tau,r)>\,$.
%
The mean field spectral weight is 
\begin{equation}
A_0(\epsilon,\vp)=\frac{2\pi}{N}\sum_\vq \,\Bigl(n_B(\bo_{\vq})+n_F(\xi_{\vp+\vq}) \Bigr)\,\delta(\epsilon+\bo_\vq-\xi_{\vp+\vq}),
\ee
where $\bo_\vq=|\vq|^2/(2 m_B)-\mu_B ,\; \xi_{\vp+\vq}=|\vp+\vq|^2/(2m_F)-\mu_F \,$.
\indent The mean-field spectral weight can  be divided into two parts. One is from  Bose factor $ n_B(\bo_\vq)$ 
\begin{equation}
A_{0{\rm b}}(\epsilon,\vp)=\frac{m_F}{2 \pi}\frac{T_0}{T}\,\frac{\sqrt{m_{B}T}}{|\vp|}\, \Phi ( \frac{|\epsilon-\xi_\vp-\mu_B|}{v_\vp \sqrt{2 m_B T}} ),
\end{equation}
 where $ \Phi (x )= \int_x^{\infty} d y \, e^{-y^2} \,$. 
The other is from  Fermi factor $n_F(\xi_{\vp+\vq})$
\begin{eqnarray}
A_{0{\rm f}}(\epsilon,\vp)=\left\{
\begin{array}{l}
0 \;\;\mbox{for} \;\; \epsilon-\mu_B\geq-\frac{(|\vp|-p_F)^2}{2\,m_B} \;,  \\ 
\frac{m_{F}}{(2\pi) |\vp|}\left[ m_B\,\tilde{\epsilon}\right]^{1/2} \;\;\mbox{for}\;\; \epsilon-\mu_B\leq-\frac{(|\vp|-p_F)^2}{2\,m_B}
\end{array}
\right.
\end{eqnarray}
, where $[\tilde{\epsilon}]^{1/2}=(\mu_{B}-\epsilon)^{1/2}- \frac{1}{m_B^{1/2}}\left|\frac{\epsilon-\mu_{B}-\xi^{f}_{p}}{v_{F}}\right|\,$. Note that $[\tilde{\epsilon}]^{1/2}=0$ when $\epsilon-\mu_B=-\frac{(|\vp|-p_F)^2}{2\,m_B}\,$.
The contribution from bose factor resembles the quasi-particle peak but it is 
severely broadened by  thermal boson with width $v_F\sqrt{m_B T}$ and it decays exponentially.  The contribution
from  fermion factor represents incoherent background which vanishes above threshold frequency\cite{lee}. \\
\indent The momentum distribution function in terms of spectral weight is given by
$n_F(\vk)=\int_{-\infty}^{\infty}\,\frac{d \omega}{2\pi}\,n_F(\omega)\,A_F(\omega,\vk) \,$.  
The momentum distribution function for electron in the mean field approximation
is \cite{yu}
\ba
n_e(\vp)&=&\sum_{\alpha}\,\int_{-\infty}^{\infty}\,\frac{d \epsilon}{2\pi}\,n_F(\epsilon)\,A_{0 \alpha}(\epsilon,\vp)  \\
&=&1-\delta+\frac{1}{N}\sum_{\vq,\sigma}\,n_{F \sigma}(\vp+\vq)\,n_B(\vq)
\ea
Since $ n_{F \sigma}(\vp+\vq) \ge 0 \,$ and $ n_B(\vq) \ge 0 \,$  it is impossible to remove $1-\delta $ states beyond  the Fermi points.  Therefore, ther is no EFS in mean-field approximation. \\
 \indent The gauge field fluctuation in decoupled approximation can be taken into account
by dressing spinon and holon Green functions. Since the spinon and holon self-energy are not gauge invariant the contribution from static gauge field diverges, thus we  consider only non-static (or inelastic ) part of gauge field.  Essentially zero-temperature result can be applied $ \Sigma_F(\epsilon) \approx (\epsilon/\chi)^{2/3}/m_F \,$ and $ \Sigma_B (\epsilon)\approx \epsilon^{3/2} m_B^{1/2}/\gamma \,$\cite{lee,ioffe}. The self-energy of boson is small compared to bare term and it can be ignored. Including  fermion self-energy  the width of the quasi-particle peak is given by $ \max[ (\epsilon/\chi)^{2/3}/m_F, v_F \sqrt{m_B T} \,] $. Thus the self-energy contributions do  not give rise to any singular effect, so that they do not change global feature of momentum distriubtion function obtained in mean field theory.
\section{Static Gauge field corrections}
In this section we outline the second order perturbation calculation of electron Green function with  static gauge field. There are five diagrams to be summed.
\be
\begin{array}{rl}
(a)&=-T^{2}\sum_{\vq,\vk,i\nu}G_B(i\nu,\vq)\,\Big[G_F(i\epsilon+i\nu,\vp+\vq)\Big]^{2}\,G_F(i\epsilon+i\nu,\vp+\vq+\vk)\, D(\vk)\,\frac{|({\bf p}+{\bf q})\times\hat{{\bf k}}|^2}{m_{F}^{2}}    \\ 
(b)&=-T^{2}\sum_{\vq,\vk,i\nu}G_B(i\nu,\vq)\,\,\Big[ G_F(i\epsilon+i\nu,\vp+\vq)\Big]^{2}\,\,\frac{D(\vk)}{2m_{F}}   \\
(c)&=-T^{2}\sum_{\vq,\vk,i\nu}G_F(i\epsilon+i\nu,\vp+\vq)\,\Big[G_B(i\nu,\vq)\Big]^{2}\,\,G_B(i\nu,\vq+\vk)\, D(\vk)\,\,\frac{|{\bf q}\times\hat{{\bf k}}|^2}{m_{B}^{2}}     \\
(d)&=-T^{2}\sum_{\vq,\vk,i\nu}G_F(i\epsilon+i\nu,\vp+\vq)\,\,\Big[G_B(i\nu,\vq)\Big]^{2}\,\,\frac{D(\vk)}{2m_{B}}   \\
(e)&=-T^{2}\sum_{\vq,\vk,i\nu}G_F(i\epsilon+i\nu,\vp+\vq)\,G_B(i\nu,\vq)\,G_F(i\epsilon+i\nu,\vp+\vq+\vk)  \\
&\times G_B(i\nu,\vq+\vk) \,D(\vk)\left ( \frac{({\bf p}+{\bf q})\cdot{\bf q}-({\bf p}+{\bf q})\cdot\hat{{\bf k}}{\bf q}\cdot\hat{{\bf k}}}{m_F m_B} \right ),
\end{array}
\ee
,where $G_B(i\nu,\vq)=[i\nu-\bo_\vq]^{-1},G_{F}(i\epsilon,\vp)=[i\epsilon-\xi_\vp]^{-1}$ are the bare boson and fermion Green functions, respectively. $(a)$ and $(b)$ are fermion self energy corrections and $(c)$ and $(d)$ are boson self-energy corrections. $(e)$ is the vertex correction.
Note that the $\vk$ integral  formally diverges logarithmically  at low $k$. \\
\indent To simplify  lengthy expressions we introduce some notations
\be
\begin{array}{rl}
z&=i\epsilon+i\nu-\xi_{\vp+\vq}=\Big[G_F(i\epsilon+i\nu,\vp+\vq) \Big]^{-1},\;\; w=i\nu-\bo_\vq =\Big[G_B(i\nu,\vq) \Big]^{-1} \\
Z&=(z-\frac{k^2}{2 m_F})^2-\vf k^2,\;\;W=(w-\frac{k^2}{2 m_B})^2-\vb k^2 \\
\Pi(z,w)&=\frac{|\vq|^2}{ m_B^2}(z-\frac{k^2}{2 m_F})^2+\frac{|\vp+\vq|^2}{m_F^2}(w-\frac{k^2}{2 m_B})^2  \\
&-2\,\frac{(\vp+\vq)\cdot \vq}{m_F m_B}\,(z-\frac{k^2}{2 m_F})(w-\frac{k^2}{2 m_B}) -k^2 \left|\frac{(\vp+\vq)\times \vq}{m_B m_F} \right|^2
\end{array}
\ee
\indent 
The first step is the integration over angle of $\vk$. For diagrams$(a)$, $(c)$ and $(e)$ the integrations give 
\be
(a)=-T^2\sum_{\vq,i\nu}G_B(i\nu,\vq)\,\Big[G_F(i\epsilon+i\nu,\vp+\vq)\Big]^{2}\,\int\frac{k\, d k}{2\pi} \frac{1}{\chi k^2} \,\frac{\frac{|\vp+\vq|^2}{m_F^2}}{z-\frac{k^2}{2m_F}+\sqrt{Z}} 
\label{fermi}
\ee
\be
(c)=-T^{2}\sum_{\vq,i\nu}\,G_F(i\epsilon+i\nu,\vp+\vq)\,\Big[G_B(i\nu,\vq)\Big]^{2}\, \int_{k=0}^{|\vp|}\frac{k\, d k}{2\pi}\frac{1}{\chi k^2}\,\frac{\frac{|\vq|^2}{m_B^2}}{w-\frac{k^2}{2m_B}+\sqrt{W}}
\label{bose}
\ee
\be
\begin{array}{rl}
(e)&=-T^2\sum_{\vq,i\nu}\,G_B(i\nu,\vq)\,G_F(i\epsilon+i\nu,\vp+\vq)\,\Lambda_{{\rm V}}(i\epsilon+i\nu,i\nu)  \\
\Lambda_{{\rm V}}&\hspace{-.1in}(i\epsilon+i\nu,i\nu)=\int\,\frac{ k dk }{2\pi}\,\frac{1}{\chi \, k^2}\frac{1}{\Pi(z,w)} \Bigg[\left|\frac{(\vp+\vq)\times \vq}{m_F m_B} \right|^2 +\frac{\,(w-\frac{k^2}{2m_B})\vf\frac{(\vp+\vq)\cdot \vq}{m_F m_B}-(z-\frac{k^2}{2m_F})\, \frac{|\vp+\vq|^2 |\vq|^2}{(m_F m_B)^2}}{\, Z^{1/2}}\,  \\
&+\frac{(z-\frac{k^2}{2m_F})\vb\frac{(\vp+\vq)\cdot \vq}{m_F m_B}-(w-\frac{k^2}{2 m_B})\frac{|\vp+\vq|^2 |\vq|^2}{(m_F m_B)^2} }{\,W^{1/2}}+  \frac{\left|\frac{\vp+\vq}{m_F}\right|^2 \left ( (z-\frac{k^2}{2 m_F})^2\,\left|\frac{\vq}{m_B}\right|^2-\frac{(\vp+\vq)\cdot\vq}{m_F m_B} (z-\frac{k^2}{2 m_F})(w-\frac{k^2}{2 m_B}) \right) }{\,\Big[Z+(z-\frac{k^2}{2 m_F})\,Z^{1/2} \Big]}  \\
&+\frac{\left|\frac{\vq}{m_B}\right|^2 \left ( (w-\frac{k^2}{2 m_B})^2\,\left|\frac{\vp+\vq}{m_B}\right|^2-\frac{(\vp+\vq)\cdot\vq}{m_F m_B} (z-\frac{k^2}{2 m_F})(w-\frac{k^2}{2 m_B}) \right) }{\,\Big[W+(w-\frac{k^2}{2 m_B})W^{1/2} \Big] } \Bigg]
\end{array}
\label{vertex}
\ee
\indent We first show that the above mentioned logarithmic divergences cancel.
The divergences ocurr in the momentum range $\, k \le \min\big(\frac{|z|m_F}{|\vp+\vq|},\frac{|w|m_B}{|\vq|} \big) \,$, and in this range the integrands of $(a)$,$(c)$ and $(e)$ can be expanded as a power series of $k^2$. For an example,
the divergent part of $(e)$ is 
$$-\frac{1}{2}\,T^2\,\sum_{\vq,i\nu} \,\Big[ G_B(i\nu,\vq)\Big]^2\,\Big[ G_F(i\epsilon+i\nu,\vp+\vq) \Big]^2\,\frac{(\vp+\vq)\cdot \vq}{m_F\,m_B}\,\int \frac{k dk}{2\pi}\,\frac{1}{\chi\,k^2} $$
Next using  the following identities it is straightforward to show that the divergent terms sum up to cancel.
\be
\begin{array}{rl}
& [G_F(i\epsilon+i\nu,\vp+\vq)]^{3}\frac{|{\bf p}+{\bf q}|^{2}}{m_{F}^{2}}=\frac{1}{2}\nabla_{{\bf p}+{\bf q}}^{2}G_F(i\epsilon+i\nu,\vp+\vq)-\frac{\big[G_F(i\epsilon+i\nu,\vp+\vq)\big]^{2}}{m_{F}}   \\
& [G_B(i\nu,\vq)]^{3}\,\frac{|{\bf q}|^{2}}{m_{B}^{2}}=\frac{1}{2}\nabla_{{\bf q}}^{2}\,G_B(i\nu,\vq)-\frac{\big[G_B(i\nu,\vq)\big]^{2}}{m_{B}}
\end{array}
\ee

\be
\begin{array}{rl}
\nabla_{{\bf p}+{\bf q}}\cdot\Bigl(\nabla_{{\bf p}+{\bf q}}G_F(i\epsilon+i\nu,\vp+\vq)\,G_B(i\nu,\vq)\Bigr)=& \\
 \Bigl(\nabla_{{\bf p}+{\bf q}}^{2}G_F(i\epsilon+i\nu,\vp+\vq)\Bigr)\,G_B(i\nu,\vq)&+\nabla_{{\bf p}+{\bf q}}G_F(i\epsilon+i\nu,\vp+\vq)\cdot\nabla_{{\bf q}}G_B(i\nu,\vq)  \\
 \nabla_{{\bf q}}\cdot\Bigl(\nabla_{{\bf q}}G_B(i\nu,\vq)\,G_F(i\epsilon+i\nu,\vp+\vq)\Bigr)&=\\
 (\nabla_{{\bf q}}^{2}G_B(i\nu,\vq))G_F(i\epsilon+i\nu,\vp+\vq)&+\nabla_{{\bf p}+{\bf q}}G_F(i\epsilon+i\nu,\vp+\vq)\cdot\nabla_{{\bf q}}G_B(i\nu,\vq)
\end{array}
\ee
 The remaining finite part is negligible in the small $|z|$,$|w|$ limit.\\
 \indent At this point it is convenient to analytically continue to real frequency representation. By summing Matsubara frequency using contour integral
the spectral weight can be expressed as
\be
\begin{array}{rl}
A(\epsilon,\vp)&=\frac{1}{N}\sum_{\vq}\,\int_{-\infty}^{\infty}\,\frac{ d x}{\pi} \Bigg[ \Big(n_B(x)+n_F(\epsilon+x) \Big)\,{\rm Re}\Big[\,G_F^R(\epsilon+x)\,G_B^A(x)\,\Lambda_{+-} \Big]   \\
&-\Big(n_B(x)+n_F(\epsilon+x) \Big)\,{\rm Re}\Big[\,G_F^R(\epsilon+x)\,G_B^R(x)\,\Lambda_{++} \Big] \Bigg], 
\end{array}
\label{spec}
\ee
where $ G_F^R(\epsilon+x)=G_F(\epsilon+x+i\delta), \;\;G_B^A(x)=G_B(x-i\delta), \, \;\; \Lambda_{\pm \pm}=\Lambda(\epsilon+x \pm i\delta,x  \pm i \delta)\,$. 
$\Lambda(\epsilon+x,x)$ consists of fermion self-energy, boson self-energy and vertex correction. 
 The mean field result corresponds to $\Lambda_{+-}=\Lambda_{++}=1$.\\
\indent When $ k  \ge \frac{|z|m_F}{|\vp+\vq|} $ or $k \ge \frac{|w|m_B}{|\vq|} $
the self energy correction and the vertex correction  become imaginary due to branch-cut singularity.
In the small $|z|$, $|w|$ limit both self-energy and vertex correction become singular. 
The singular contribution from  fermion self energy $(a)$ is of the form
\be
\Lambda_F=\pm i \, \frac{|\vp+\vq|^2}{m_F^2}\, \,\big[G_F (\epsilon+x)\big]^2,
\ee
where the sign is plus if Green function is retarded and vice versa.
Similarly from boson self energy $(c)$
\be
\Lambda_B=\pm i \, \frac{|\vq|^2}{m_B^2}\, \,\big[G_B (x)\big]^2
\ee
By examining Eq.(\ref{spec}) it is clear that $\Lambda_F$ is divergent at the pole of ${\rm Im}\,G_B(x)$ and $\Lambda_B$ is divergent at the pole of ${\rm Im}\,
G_F(\epsilon+x)$. In the vertex correction Eq.(\ref{vertex}), the second and the third term give dominant imaginary contributions. At the pole of ${\rm Im}\,G_B(x)$ the third term  of Eq.(\ref{vertex})  cancels out between the first and  the second term of $A(\epsilon,\vp)$ Eq.(\ref{spec}) and  the leading divergent  $-z\frac{|\vp+\vq|^2|\vq|^2}{(m_F m_B)^2} Z^{-1/2}$ part of the second term of  Eq.(\ref{vertex}) cancels the fermion self-energy term $\Lambda_F$. 
Analogously, the singular boson self-energy term is also cancelled by the third term of vertex correction at the pole of ${\rm Im}\,
G_F(\epsilon+x)$.\\
\indent Therefore, the leading contribution to $\Lambda(\epsilon+x,x)$ is given by
\be
\begin{array}{lr}
\Lambda(\epsilon+x,x)&=\int \frac{ k dk}{2\pi}\frac{1}{\chi\, k^2 \Pi(z,w)}\Bigg[ \frac{\frac{k^2}{2m_F}\, \frac{|\vp+\vq|^2 |\vq|^2}{(m_F m_B)^2}-\frac{k^2}{2m_B}\vf\frac{(\vp+\vq)\cdot \vq}{m_F m_B}}{\, Z^{1/2}} \\
&+\frac{\frac{k^2}{2 m_B}\frac{|\vp+\vq|^2 |\vq|^2}{(m_F m_B)^2}-\frac{k^2}{2m_F}\vb\frac{(\vp+\vq)\cdot \vq}{m_F m_B} }{\,W^{1/2}}\,\Bigg]
\end{array}
\label{domi}
\ee
The integrand becomes large when $w=0$ or equivalently at the pole of ${\rm Im}\,G_B(x,\vq)\,$.
The integration over $k$ can be done and the correction to the spectral weight
can be written as

\be
A_1(\epsilon,\vp)\approx-\frac{v_F T}{\chi}\int\frac{d^2 \vq}{(2\pi)^2}\frac{{\rm sgn}(z)\,\cos \phi\Big[n_F(\epsilon+\bo_\vq)+n_B(\bo_\vq) \Big]}{z^2 \big(|z|m_F |\vq|/|\vp|+z m_F \cos \phi|\vq|/|\vp|-|\vq|^2\sin^2 \phi\big)^{1/2}}, 
\ee
where $\phi$ is the angle between $\vp$ and $\vq$. In angular integral major contributions come from near $\phi=0$ and $\phi=\pi$. Performing $\vq$ integration we get
\be
A_1(\epsilon,\vp)\approx-\frac{T}{\chi}\frac{1}{|\epsilon-\mu_B-\xi_\vp|}\Big[{\rm const.}+\Phi(\frac{|\epsilon-\mu_B-\xi_\vp|}{v_\vp \sqrt{2 m_B T}}\,)\Big]
\ee
By comparing with the mean field spectral weight, $ A(\epsilon,\vp)=A_0(\epsilon,\vp)+A_1(\epsilon,\vp) \,$ can be written as
\be
A(\epsilon,\vp)=A_{0 b}(\epsilon,\vp)\, \Big(1-\frac{\tau_{{\rm in}}^{-3/2}}{|\epsilon-\mu_B-\xi_\vp| \sqrt{T}}\Big)+A_{0 f}(\epsilon,\vp)\,\Big(1-\frac{\tau_{{\rm in}}^{-3/2}}{|\epsilon-\mu_B-\xi_\vp| \sqrt{\tilde{\epsilon}}} \Big),
\label{A}
\ee
which is the main result of this paper.

\section{Discussions}
 First, $\epsilon^{-3/2}$ singularity of Eq.(\ref{A}) can be understood by explicitly gauge invariant path integral method.
Neglecting statistics of spinon and holon the electron Green function can be expressed as \cite{boris,mirlin}
\be
\begin{array}{rl}
G^R(r,t)&=\int_{r_F(0)=0}^{r_F(t)=r}\big[D r_F(t^{\prime}) \big]\,\int_{r_B(0)=0}^{r_B(t)=r}\big[D r_B(t^{\prime}) \big]\,\exp\Big[i\int_0^t d t^{\prime}\,\frac{m_F}{2}\left(\frac{d r_F}{d t^{\prime}} \right)^2 \\
&-i\int_0^t d t^{\prime} \frac{m_B}{2}\left(\frac{d r_B}{d t^{\prime}} \right)^2-\frac{T}{2\chi} \,S(r_F,r_B) \Big],
\end{array}
\ee
where $S(r_F,r_B)$ is the non-oriented area enclosed by fermion path $r_B(t^{\prime})$ and boson path $r_B(t^{\prime})$.
In semi-classical approximation the area term can be written as
\be
S(r_F,r_B)\approx \frac{v_F T}{2\chi} \int_0^t d t^{\prime}  \,|r^{\perp}_F(t^{\prime})-r^{\perp}_B(t^{\prime})|,
\ee
where $r^{\perp}_F$,$r^{\perp}_B$ denote the component of fermion and boson path
which are perpendicular to the straight line connecting $0$ and $r$. By straightforward perturbative calculation the first order term can be shown to be proportional to $t^{3/2}$, which in turn implies $\epsilon^{-3/2}$ singularity.
Note that $t^{1/2}$ factor originates from the fact that the {\em transverse} displacements scale like $\sqrt{t}$.\\
\indent From Eq.( \ref{A} ) it is clear that the perturbation theory breaks down  at low energy. Thus the resummation or other non-perturbative methods are required.
If the ladder diagrams are summed  the corrections can be summed like a geometric series. After the summation the low energy singularity disappears and the gauge field correction does not influence the momentum distribution function. \\
\indent If  the maximally crossed diagrams are summed the corrections sum up exponentially, so that the gauge field correction becomes exponentially small at low energy. This result agrees with path integral interpretation because the low energy  corresponds to large area in path integral approach and the large area are suppressed exponentially. However, all of the above resummations are not controlled approximations and the results obtained are not reliable.  \\
\indent For one-particle problem the semi-classical expansion can provide some non-perturbative informations \cite{boris}. 
 For a fast particle in random static magnetic field the gauge invariant one-particle Green function can be obtained using semi-classical expansion. The Green function is characterized by phase coherence time $\tau^{-1}_{\phi}=\Big((\frac{T}{\chi})^2 \,\frac{r^2}{ m t^2} \Big)^{1/3} $ and a series of branch cut singularities.  The above two features can not be observed in perturbative expansion. If the mass is set equal to boson mass $m=m_B$ and if  $r/t $  is taken as Fermi velocity  $\tau^{-1}_{\phi}$ becomes identical with  $\tau^{-1}_{{\rm in}}$. This  fact implies that some features of the one-particle Green function will appear in our two-particle electron Green function . We leave the semi-classical approach to the two particle electron Green function for future study.


\vskip 0.5cm
\centerline{\bf ACKNOWLEDGEMENTS}
 This work was supported in part by the National Science Foundation (DMR 91-20000) through the Science and Technology Center for Superconductivity.



\begin{references}
\bibitem{shen} Z. X. Shen, {\it et al.},\,
Phys.  Rev.  Lett.  {\bf 64}, 2442 (1990).
\bibitem{olson} C. G. Olson, {\it et al.},\,
Phys. Rev. B {\bf 42}, 381 (1990).
\bibitem{larkin} L. B. Ioffe and A. Larkin, Phys. Rev. B {\bf 39}, 8938 (1989).
\bibitem{lee} N. Nagaosa and P. A. Lee,\,
Phys.  Rev.  Lett. {\bf 64}, 2450 (1990) ; P. A. Lee and N. Nagaosa, Phys. Rev. B {\bf 46}, 5621 (1992).
\bibitem{paul} L. B. Ioffe and P. B. Wiegmann, Phys. Rev. Lett. {\bf 65}, 1653 (1990).
\bibitem{ioffe} L. B. Ioffe and G. Kotliar, \,
Phys.  Rev.  B {\bf 42}, 10348 (1990).

\bibitem{yu} S. Feng, J. B. Wu, Z. B. Su and L. Yu, \,
Phys.  Rev. B  {\bf 47}, 15192 (1993).
\bibitem{boris} B. L. Altshuler and L. B. Ioffe, \, 
Phys. Rev. Lett. {\bf 69}, 2979 (1992)
\bibitem{mirlin} A. D. Mirlin, E. Altshuler, and P. W\"{o}lfle, \,
Preprint, cond-mat/9507081


\end{references}
\end{document}